# Broadband and Ultra-Compact Adiabatic Coupler Based on Linearly Tapered Silicon Waveguides

Ahmed Bayoumi[1,2], Ahmed Khalil[1,3], Pol Van Dorpe[1,3], Maumita Chakrabarti[1], Dimitrios Velenis[1], Philippe Absil[1], Filippo Ferraro[1], Yoojin Ban[1], Joris Van Campenhout[1], Wim Bogaerts[2,1,*], Qingzhong Deng[1,*]

[1] imec, Kapeldreef 75, 3001 Leuven, Belgium
[2] Photonics Research Group, Department of Information Technology, Ghent University, Ghent, Belgium
[3] Department of Physics and Astronomy, KU Leuven, Leuven, Belgium
*Corresponding authors: qingzhong.deng@imec.be, wim.bogaerts@ugent.be

**Abstract** *2×2 couplers are crucial in many photonics applications but are often wavelength sensitive. Alternative designs could be large, complex, or lossy. We present a broadband ultra-compact adiabatic coupler with minimal coupling variation and the least silicon taper length of $1.44\ \mu m$, to our best knowledge. © 2025 The Author(s)*

## 1. Introduction

2×2 optical couplers play a crucial role in photonic integrated circuits (PICs) applications such as in wavelength division multiplexing (WDM) [1,2], modulation [3], and signal switching [4]. Symmetric straight directional couplers (DCs) are often used as the 2×2 couplers. However, they are quite wavelength sensitive, particularly in high-index contrast systems, which often cause performance degradation for many PIC applications. Ideally, 2×2 couplers should have a broad wavelength operation range, have a compact footprint, be capable of broadband coupling at arbitrary coupling ratios, and attain low loss. Several approaches have been proposed to achieve highly performing 2×2 couplers including multimode interferometers [5], DCs with phase compensation sections [6], subwavelength grating based DCs [7], rib waveguide based DCs [8], adiabatic couplers [9], and bent DCs [10]. In these studies, several issues exist. For instance, relatively high insertion loss [5-9] which can be particularly problematic for dense WDM or switching applications where several 2×2 couplers are used causing loss accumulation. Moreover, some designs exhibit a relatively large footprint [8, 10] or have fabrication complexities [7].

Adiabatic coupler is another promising 2×2 optical coupler scheme that commonly consists of two adjacent linearly tapered waveguides. The core principle involves gradually tapering the waveguide width to maintain the power in one optical system mode [11,12]. The key design parameters are the taper length, the waveguide width difference at the taper start and end. The waveguide width difference at the taper end is exploited to manipulate the coupling ratio [13-14]. The other two parameters are often chosen intuitively to be sufficient to ensure adiabaticity, without attention of their influence to the wavelength dependency of the coupling ratio. This results in non-optimized wavelength dependence and usually a large size in the reported adiabatic couplers.

In this work, we show that the wavelength sensitivity of the coupling ratio in adiabatic couplers could be significantly reduced by solving a minmax optimization problem with the taper length and the waveguide width difference at the taper start as the free parameters. This also allows for less asymmetry in the design that enables the mode to evolute fast, reducing the footprint dramatically. Based on that approach, we demonstrate numerically a 3-dB adiabatic coupler with a significantly low coupling variation of 0.02 over 120 nm wavelength range covering the O-band. This represents, to our best knowledge, the least coupling variation for silicon adiabatic couplers over the presented wavelength range, that are compatible with standard silicon photonics fabrication platforms. Compared to the traditional straight DC counterpart, this work has 16.6x reduction in coupling variation. Further, the optimization results in an ultra-compact taper length of $1.44\ \mu m$, which is the shortest mode evolution length for silicon 3-dB couplers reported in the literature to our best knowledge. Moreover, novel low loss third order polynomial interconnected circular (TOPIC) bends that attain continuous curvature and curvature derivative [1] were exploited in the design to achieve an insertion loss of ≤0.03 dB over the presented 120 nm wavelength range. Overall, the presented device is broadband, ultra-compact, can enable arbitrary coupling ratios, and is low-loss, rendering it a highly competitive 2×2 coupler to be used in next-gen PICs applications.

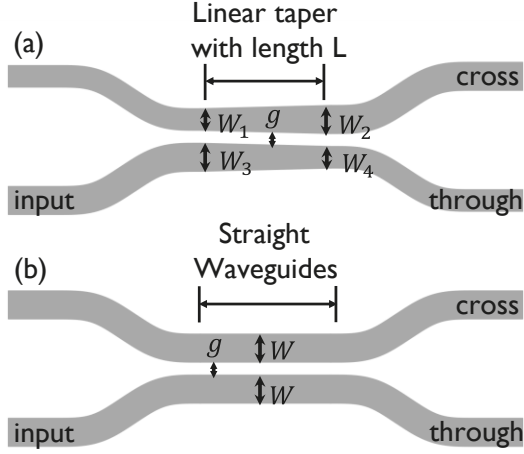

**Figure 1** Schematic of the proposed tapered DC (a), where the top waveguide width increases in the propagation direction, while the bottom waveguide width decreases. The schematic of the traditional straight DC (b) with fixed waveguide width of $W$. Both designs have a fixed gap of $g$. The schematic is not in scale.

## 2. Tapered DC Design and Results

The proposed adiabatic coupler consists of two linearly tapered waveguides whose width variations have opposite signs, with a taper length of $L$, which are connected to low-loss TOPIC S-bends at the input/output sections (Fig. 1(a)). The traditional DC counterpart is shown in Fig. 1(b) for benchmarking. The adiabatic coupler waveguides widths can be described as in Eq (1-2).

$$W_2 = W_1 + \Delta W * t_p \quad (1),$$
$$W_4 = W_3 - \Delta W * t_p \quad (2),$$

where the top and bottom waveguides are linear tapers with initial and final waveguide widths of $(W_1, W_2)$ and $(W_3, W_4)$, respectively. $\Delta W = W_3 - W_1$ is the waveguide width difference at the taper start. $t_p \in [0,1]$ is the taper percentage ratio, that controls waveguide width difference at the taper end. $t_p$ should be set as the required cross coupling ratio (i.e. $t_p = 0.5$ for a 3-dB coupler).

The complementary tapering in the top and bottom waveguides facilitates efficient energy transfer between the waveguides because the optical mode preferentially remains confined in the region with the higher effective refractive index. Further, the gradual variation in dimensions minimizes scattering and reflection losses, ensuring robust mode conversion and improved coupling efficiency between the waveguides [15].

As depicted by coupled mode theory, the narrower waveguide width has less mode confinement with a larger coupling strength allowing a shorter coupler length [15]. $W_1$ is chosen to be $0.26\ \mu m$ to achieve higher coupling strength without introducing additional excess loss according to 3D FDTD simulations. To achieve broadband coupling at a specific ratio, $\Delta W$ and $L$ must be optimized to minimize the coupling deviation from the desired coupling ratio over a broad wavelength range. This can be formulated as a minmax optimization problem,

$$\min_{\Delta W, L} \left( \max_{\lambda \in [\lambda_1, \lambda_2]} |P_c(\lambda) - P_{c,desired}| \right) \quad (3),$$

where $\lambda_1 = 1.25\ \mu m$, $\lambda_2 = 1.37\ \mu m$, $P_c(\lambda)$ is the cross coupled power as a function of wavelength, and $P_{c,desired}$ is the desired cross coupled power value. By solving the optimization problem in Eq. (3), a $(\Delta w, L)$ pair can be determined that minimizes the coupling variation. For instance, for the 3-dB coupler, the coupling variation could be minimized to merely 0.02 over the presented wavelength range with $\Delta w = 0.06\ \mu m$ and an ultra-compact taper length of $L = 1.44\ \mu m$, as shown in Fig. 2. The coupling spectrum of the proposed 3-dB splitter is shown in Fig. 3(a) with 16.6x reduction in the coupling variation as compared to the traditional straight DC in Fig. 3(b).

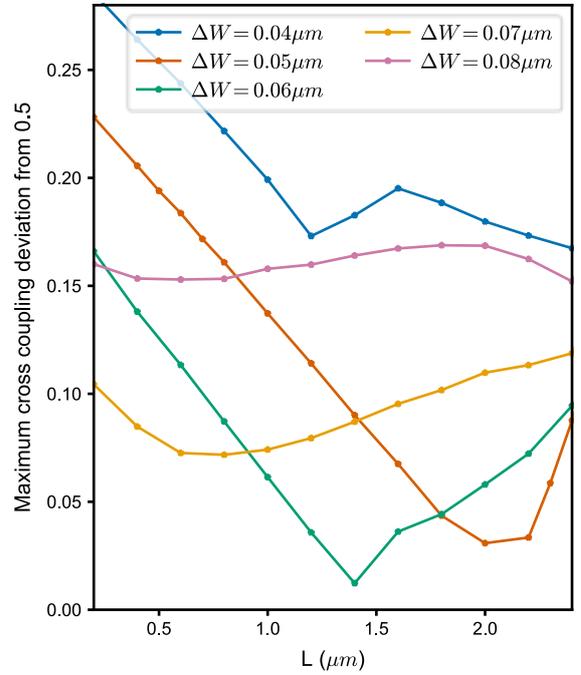

**Figure 2** Maximum cross coupling deviation from 0.5 over 120 nm wavelength range as a function of the taper length ($L$) and the initial waveguides width difference ($\Delta W$). The fixed design parameters are $W_1 = 0.26\ \mu m, g = 0.08\ \mu m, t_p = 0.5$. Each line represents the coupling deviation at a certain $\Delta W$. The coupling deviation is minimal at $\Delta W = 0.06\ \mu m$ with a compact $L = 1.44\ \mu m$.

To illustrate the adiabatic nature of the proposed coupler, the power confinement ratio in the TE even and odd modes were extracted at the middle of the taper region for the proposed 3-dB coupler. Figure 4(a) shows the $Hy$ field plot representing the TE mode of the coupler, where ~98.3 % of the power remains in the TE even mode as shown in Fig. 4(b) and ~1.7% of the power is in the TE odd mode at $\lambda = 1.31\ \mu m$, which shows the smooth mode evolution taking place in this design and confirms the adiabatic nature of the tapering.

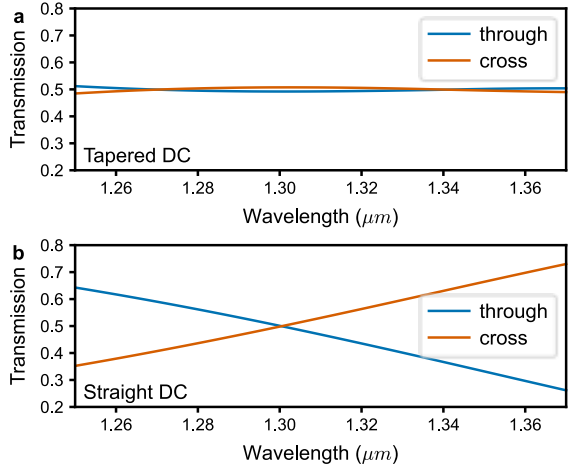

Figure 3 Optical transmission of the 3-dB coupler using the proposed adiabatic coupler (a) as compared to the traditional DC (b) over 120 nm wavelength range covering the O-band. Significant reduction of 16.6x in the coupling variation is achieved using the proposed design compared to the straight DC counterpart.

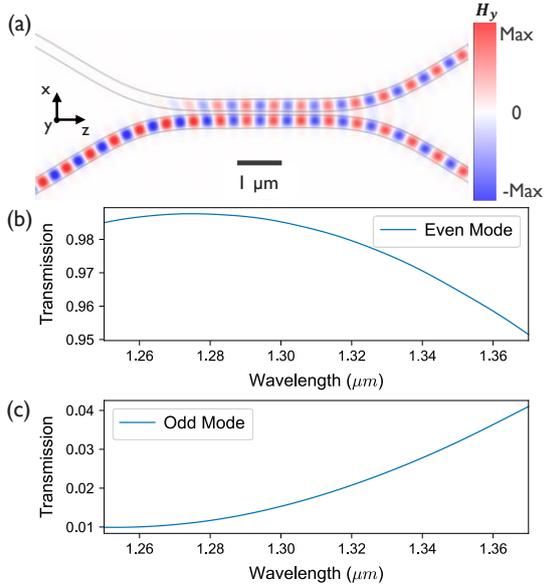

Figure 4 $H_y$ field plot of the proposed 3-dB coupler at $\lambda = 1.31\,\mu m$ (a). TE even (b) and odd (c) mode power expansions at the middle of the coupler showing most of the power (98.3% at $\lambda = 1.31\,\mu m$) existing in the TE even mode, proving the adiabatic nature of the coupler.

$t_p$ facilitates the mode evolution to achieve the corresponding splitting ratio, for instance 3-dB splitting can be achieved by maintaining equal output waveguide width. Further, $L$ could affect the amount of coupling that takes place and has a direct control on the cross coupled value. Finally, $\Delta W$ has a large impact on the asymmetry of the design, where a larger $\Delta W$ value allows for lower cross coupling values, while smaller $\Delta W$ values allow for larger cross coupling values. Nonetheless, with a fixed $\Delta W =$ 0.06 $\mu m$, broadband coupling can take place at coupling ratios of 0.2, 0.3, 0.4, 0.5, and 0.55 as presented in Fig. 5(a), where $W_1 = 0.26\,\mu m$ and $t_p$ equals the desired cross coupling ratio. Furthermore, it is shown that for broadband coupling, there exists a quadratic relationship between the taper length, $L$, and the taper percentage, $t_p$ (and the cross coupling value), as shown in Fig. 5(b) for the presented coupling ratios, at a fixed $\Delta W$, where a larger taper length is required to achieve larger cross coupling ratio as expected.

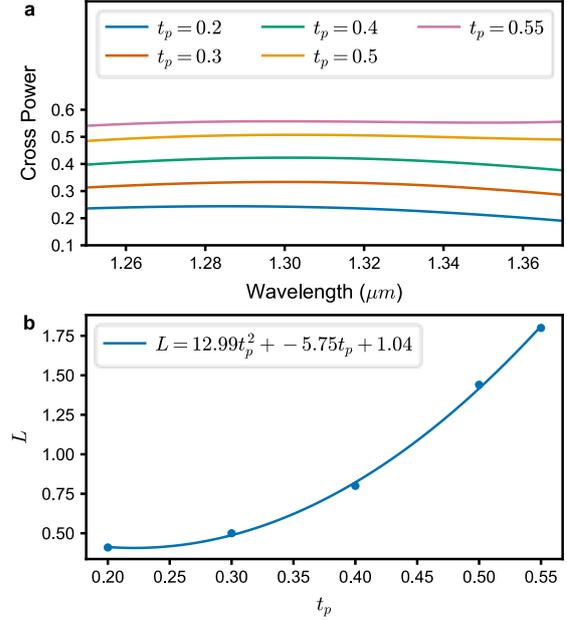

Figure 5 Demonstration of broadband coupling using the proposed tapered DC for cross coupling values of 0.2, 0.3, 0.4, 0.5, and 0.55 over 120 nm wavelength range (a). Taper length ($L$) as a function of the taper percentage ($t_P$) (b). $t_P$ is designed to have the same value as the required cross coupling value. Dots represent the data while the line is quadratic fitting. All the presented devices have the same parameters except for $t_P$ and $L$.

### 3. Conclusions

We have shown that by solving a minmax optimization problem for the coupling deviation of adiabatic couplers, the coupling variation could be significantly minimized by optimizing the taper length and the waveguide width difference at the taper start. Based on this approach, we demonstrated numerically a 3-dB adiabatic coupler with a coupling variation of 0.02 over 120 nm wavelength range covering the O-band, achieving 16.6x reduction in the coupling variation as compared to the traditional straight DC counterpart. Furthermore, the proposed 3-dB coupler has an insertion loss of ≤0.03 dB, and an ultra-compact taper length of $1.44\,\mu m$ that is the shortest length reported to our best knowledge. Arbitrary coupling ratios are achieved by adjusting the final tapers waveguides width. Overall, the proposed coupler and design scheme can enable optimum 2×2 couplers in next-gen photonics applications.

## Acknowledgements

This work was supported by imec's industry-affiliation R&D program "Optical I/O". A. K. acknowledges the FWO scholarship fund (1SHDN24N).